\documentclass[preprint,final,1p,article]{elsarticle}
\usepackage{lineno}
\modulolinenumbers[2]
\journal{Nuclear Physics A}
\usepackage{amssymb}
\usepackage{graphicx}
\usepackage{color}
\usepackage{amsthm}
\usepackage{amsmath}
\usepackage{amssymb}
\usepackage[colorlinks,citecolor=blue,linkcolor=red,anchorcolor=blue,filecolor=blue,urlcolor=blue]{hyperref}
\begin{document}
\begin{frontmatter}
\title{Exploring the ground state bulk and decay properties of the nuclei in superheavy island \tnoteref{Decay properties of Superheavy Island}}
\author[1]{Nishu Jain}
\ead{nishujain1003@gmail.com}
\author[1]{Raj Kumar}
\ead{rajkumar@thapar.edu}
\author[2,3]{M. Bhuyan\corref{2,3}}
\cortext[2,3]{Corresponding author}
\ead{bunuphy@um.edu.my}
\address[1]{School of Physics and Materials Science, Thapar Institute of Engineering and Technology, Patiala 147004, India}
\address[2]{Center for Theoretical and Computational Physics, Department of Physics, Faculty of Science, University of Malaya, Kuala Lumpur 50603, Malaysia}
\address[3]{Institute of Research and Development, Duy Tan University, Da Nang 550000, Vietnam}

\begin{abstract}
The $\alpha$-decay half-lives of 204 superheavy nuclei covering the range $114 \leq Z \leq 126$ have been investigated using the relativistic mean-field model (RMF) for NL3$^*$ parameter set.  The ground state bulk properties such as binding energy, quadrupole deformation parameter ($\beta_2$ ), and root-mean-square charge radii for these nuclei are analyzed. Four different semi-empirical formulae, namely, the universal decay law (UDL), the Viola-Seaborg (VSS) formula, the modified universal decay law (MUDL), and the modified Brown formula (MBrown), are used to obtain the $\alpha$-decay half-lives for the considered nuclei. To examine the applicability of relativistic mean-field model within NL3$^*$ parametrization, the $\alpha$-decay energies, and the half-lives of a few known superheavy nuclei within the range 102 $\leq$  Z $\leq$ 118 are calculated and the results are compared with the experimental data along with the theoretical predictions.  The $\alpha$-decay energies ($Q$-values) are estimated from the binding energies of the parent, and daughter from the RMF (NL3$^*$) parameter set. The calculated results are compared with macroscopic-microscopic Finite-Range-Droplet-Model (FRDM), Global Nuclear Mass Model (WS3, WS4), Weizsacker-Skyrme mass model (WS$^*$) predictions, and the experimental data, wherever available.  The possible standard deviations are also estimated for experimental and various theoretical predictions. We find a good consistency for the experimental-to-UDL, FRDM-to-UDL, and WS4-to-UDL estimates of the decay energy and corresponding half-life. The present analysis provides the theoretical predictions within the microscopic model for the upcoming experiments on the superheavy region.
\end{abstract}
\begin{keyword}
\texttt{Relativistic Mean Field; Alpha decay; Q-value; half-lives; Analytical formula; Superheavy Nuclei}
\end{keyword}
\end{frontmatter}
\section{Introduction}
\label{intro}
The investigation of unknown superheavy elements (SHEs) and their synthesis has become a fascinating area in modern-day nuclear physics from both theoretical and experimental aspects \cite{Hofm98,Hami13,gan04,ogan13}. The $\alpha$-decay energy $Q_{\alpha}$ and the corresponding half-lives are used as a probe to study the island of stability in the neutron-rich heavy and superheavy mass region. Alpha decay is one of the most powerful tools to analyze the existence of heavy and superheavy nuclei (SHN) \cite{geig11,gamo28,cond28,moll97,chow08}. The alpha decay phenomenon has been successfully explained as a quantum tunneling process. The presence of the quantum shell effect in the SHN is the only reason for their stability. If the nuclear stability is increased around the magic number, the location of nuclear stability can be confirmed. In other words, the quantum shell effect stabilizes the superheavy nucleus against the Coulomb repulsion. Experimentally, $\alpha$-decay is used to investigate the new elements through subsequent $\alpha$-decay chains from unknown to a known region of the nuclear chart. Recently, the interest in the $\alpha$-decay nuclide has been revived in the highly neutron-to-proton asymmetry systems, including drip-lines due to the evolution of radioactive ion beams and new detector technology under low temperature. Elaborately, the large neutron-to-proton asymmetry in the SHN can be used to explore deeper information about the nuclear structure, which is essential for understanding the certain concepts such as stability islands, magic numbers, deformed nuclei, spin and parity, nuclear clustering, and nuclear interaction. \\ \\
Various theoretical models such as density-dependent M3Y (DD-M3Y) effective interaction \cite{sama07}, the cluster decay model \cite{xu06}, the Coulomb and proximity potential model \cite{sant98}, the super-symmetric fission model \cite{poen06,bhat08}, the generalized liquid drop model (GLDM) \cite{Roye24,Dong10}, the unified model for $\alpha$-decay and alpha capture (UMADAC Method) \cite{deni05,deni10}, the coupled channel approach \cite{deli06,pelt07}, and the universal curve for alpha and cluster radioactivity in a fission theory \cite{poen91} have been used to express the $\alpha$-decay process. In parallel to these models, many theoreticians presented empirical and semi-empirical formulae with the help of which experimentalists can find precise values of $\alpha$-decay half-lives for the superheavy nuclei (SHN) \cite{sant98,poen06,Qian12,Akra18}. In the present scenario, the superheavy nuclei are synthesized by using cold, and hot fusion reactions. In the cold fusion reactions \cite{Armb85}, SHN has been synthesized from $Z$ = 110 to $Z$ = 112. Also, the SHN with $Z$ = 113 – 118 has been synthesized by hot fusion reactions \cite{ogan15}. Although a few superheavy isotopes and isomers with $Z \le 118$ have been discovered, still most of their properties like spin-parity, shape degrees of freedom, and stability remain unknown at present. \\ \\  
In the search of new isotopes of superheavy nuclei, the location of shell gap plays an important role. The stability decreases as the gap reaches from spherical magicity at $Z$ = 82, $N$ = 126 to $Z$ = 114, $N$ = 178 - 184 \cite{saxe19}. But to locate the shell gap in transuranium region (deformed in shape)  is a challenging task and requires accurate calculations. In the study of the superheavy region, it was predicted that the most stable configuration is at $N$ = 162 \cite{cwio94}. In another study, the stable configuration was predicted at the doubly magic nucleus with $Z$ = 108 and $N$ = 162 \cite{paty91}. Earlier studies also predict the doubly magic nucleus with $Z$ = 114 and $N$ = 184 \cite{sobi66}. These results are also confirmed in other investigations \cite{meld67,hofm19}. In later studies, for example, the Yukawa plus exponential model (YPE) with Wood-Saxon single-particle potential \cite{munt03,munt03a}, the next predicted spherical doubly magic number was $^{298}114$. According to Skyrme-Hartree-Fock-Bogoliubov method (Non-relativistic Model), it is predicted that the next expected magic number will be at $Z$ = 120 \cite{cwio99}. The relativistic mean- field formalism provides an exact microscopic description of nuclei from every part of the nuclear chart \cite{giul19}. The microscopic calculations predicted the various region of stability like $Z$ = 114, $N$ = 184 \cite{rutz97}; $Z$ = 120, $N$ = 172 \cite{rutz97,gupt97,patr99,krup00}; $Z$ = 120, $N$ = 184 \cite{gupt97,patr99,bhuy12}; and $Z$ = 124 or 126, $N$ = 184 \cite{rutz97,krup00,cwio05}. Also in the framework of relativistic continuum Hartree-Bogoliubov theory, the magic proton number at $Z$ = 120, 132, 138 and magic neutron number at $N$ = 172, 184, 198, 228, 238, 258 \cite{zhan05} are predicted. Recently, Taninah \textit{et al.} \cite{Tani20} systematically investigated the bulk nuclear and fission properties of even-even actinides and superheavy nuclei with Z = 90-120. Using the covariant energy density formalism for DD-PC1, DD-ME2, NL3$^*$, and PC-PK1 parameter sets, possible shell closure at $Z = 120$, $N = 184, 258$ are predicted based on the nuclear bulk and decay properties \cite{Tani20}. \\ \\
The nuclear models that are mainly used to calculate the binding energy are macro-microscopic, viz., Bethe-Weizsacker mass formula \cite{heyd99,wesl68}, Finite-range liquid drop model (FRDM) \cite{moll97,moll95}, and in the microscopic models includes the Relativistic mean-field model \cite{Bogu77,Sero86,Ring96}, the Skyrme-Hartree-Fock \cite{skyr56,dech80}, and the finite-range Gogny interaction \cite{gonz17,gonz18}. Various theoretical observations and fundamental efforts have been applied to reach the magic islands and synthesize the SHN beyond $Z$ = 114 with long alpha decay half-lives. It is a challenging issue to understand the structural properties of the nuclei in the superheavy mass region and more helpful for understanding the concept of the ``Island of stability" beyond the spherical doubly magic nucleus. The theoretical model that reproduces the experimental data for the structure and decay properties at the microscopic level, can be used to predict the $\alpha$-decay half-lives of those nuclei for which their experimental data have not been reported yet \cite{Bisw21}. Here our main aim is to determine the bulk and decay properties of the superheavy island with a microscopic model. The extensive analysis includes the $Q_\alpha$ values and $\alpha$-decay half-lives of $even-even$ superheavy nuclei with $114\le Z \le 126$ and $282\le A \le 350$. The relativistic mean-field with NL3$^*$ parameter sets \cite{lala09}, which is the refitted version of NL3 force parameter \cite{lala97} is employed for the present investigations. The Finite drop liquid model (FRDM), Global Nuclear Mass Model (WS4), and the available experimental data are taken for comparison. Four different semi-empirical formulae, i.e., the universal decay law (UDL) \cite{manj19}, Viola–Seaborg semi-empirical formula (VSS) \cite{manj19}, modified the universal decay law (MUDL) \cite{akra19}, and Modified Brown formula (MBrown) \cite{isma20} are used to obtain the $\alpha$-decay half-lives and their relative predictivity in the superheavy region. The possible mean or standard deviations are calculated with respect to the available experimental data to determine the applicability of the model for the superheavy island. \\ \\
This paper is organized as follows: a brief description of the relativistic mean-field (RMF) formalism is presented in Sec. \ref{theory}. The results of relativistic mean-field calculations for nuclear structure and decay properties are shown in Sec. \ref{result}. Sec. \ref{summary} concludes the brief results and discussions of the present work.

\section{Relativistic Mean-Field Formalism}
\label{theory} \noindent
The mean-field treatment of QHD has been widely used to describe the nuclear structure and infinite nuclear matter characteristics \cite{Bogu77,Sero86,Bhuy11,Bhuy15,lala99,Type99,Niks02}. In the relativistic mean-field theory, the nucleus is supposed as a combined system of nucleons (neutron and proton) interacting through the exchange of mesons and photons \cite{Sero86,Ring96,Paar07,Rein89,Vret05,Meng06,Bend03}. The contributions from the meson fields are expressed either by mean fields or by point-like interactions between the nucleons \cite{Niko92,Burv02}, and to reflect the correct saturation properties of infinite nuclear matter, the non-linear coupling terms \cite{Bogu77,Broc92} or the density depending coupling constants \cite{Type99,Niks02,Niks08,Fuch95,Hofm01} are introduced. The relativistic Lagrangian density for a nucleon-meson many-body system \cite{Bogu77,Sero86,Ring96,Bhuy11,Bhuy15,lala99,Paar07,Rein89,Vret05,Meng06,Bend03,Bhuy18,Bhuy20,Patr09,Niks11,Zhao12} is as following:
\begin{eqnarray}
{\cal L}&=&\overline{\psi}\{i\gamma^{\mu}\partial_{\mu}-M\}\psi +{\frac12}\partial^{\mu}\sigma
\partial_{\mu}\sigma \nonumber \\
&& -{\frac12}m_{\sigma}^{2}\sigma^{2}-{\frac13}g_{2}\sigma^{3} -{\frac14}g_{3}\sigma^{4}
-g_{s}\overline{\psi}\psi\sigma \nonumber \\
&& -{\frac14}\Omega^{\mu\nu}\Omega_{\mu\nu}+{\frac12}m_{w}^{2}\omega^{\mu}\omega_{\mu}
-g_{w}\overline\psi\gamma^{\mu}\psi\omega_{\mu} \nonumber \\
&&-{\frac14}\vec{B}^{\mu\nu}.\vec{B}_{\mu\nu}+\frac{1}{2}m_{\rho}^2
\vec{\rho}^{\mu}.\vec{\rho}_{\mu} -g_{\rho}\overline{\psi}\gamma^{\mu}
\vec{\tau}\psi\cdot\vec{\rho}^{\mu}\nonumber \\
&&-{\frac14}F^{\mu\nu}F_{\mu\nu}-e\overline{\psi} \gamma^{\mu}
\frac{\left(1-\tau_{3}\right)}{2}\psi A_{\mu}.
\label{lag}
\end{eqnarray}
Here $\psi$ represents the Dirac spinors of the nucleons. and $g_\sigma(m_\sigma)$, $g_\omega(m_\omega)$, and $g_\rho(m_\rho)$ are the coupling constants for $\sigma$, $\omega$, and $\rho$ mesons, respectively. $\tau$ and $\tau_3$ represents the isospin and third component of isospin. The constants $g_2$, $g_3$, and $e^2/4\pi$ are the coupling constants for self-interacting non-linear $\sigma$-meson field and photon, respectively. $A_\mu$ and M represents electromagnetic field and the nucleon mass, respectively. And the vector field tensors for the $\omega^\mu$, $\vec{\rho}_{\mu}$ and photon fields are given by  
\begin{eqnarray}
F^{\mu\nu} = \partial_{\mu} A_{\nu} - \partial_{\nu} A_{\mu} \nonumber \\
\Omega_{\mu\nu} = \partial_{\mu} \omega_{\nu} - \partial_{\nu} \omega_{\mu} \nonumber \\
\vec{B}^{\mu\nu} = \partial_{\mu} \vec{\rho}_{\nu} - \partial_{\nu} \vec{\rho}_{\mu}.
\end{eqnarray}
respectively.
Here, the fields for the $\sigma$-, $\omega$- and isovector $\rho$- meson is denoted by $\sigma$, $\omega_{\mu}$, and $\vec{\rho}_{\mu}$, respectively. The electromagnetic field is defined by $A_{\mu}$. The quantities, $\Omega^{\mu\nu}$, $\vec{B}_{\mu\nu}$, and $F^{\mu\nu}$ are the field tensors for the $\omega^{\mu}$, $\vec{\rho}_{\mu}$ and photon fields, respectively.\\ \\
From the above Lagrangian density, we obtain the field equations for the nucleons and the mesons. These field equations are solved by developing the upper and lower components of the Dirac spinors and the boson fields in an axially deformed harmonic oscillator basis, with an initial deformation $\beta_{0}$. The set of coupled equations is solved numerically by a self-consistent iteration method. The centre-of-mass motion energy correction can be calculated by the conventional harmonic oscillator formula \cite{Long04} given as
\begin{equation}
E_{c.m.}=\frac{3}{4}(41A^{-1/3}),
\end{equation}
where A is the mass number. The total quadrupole deformation parameter $\beta_2$ is calculated from the resulting proton and neutron quadrupole moments, as
\begin{equation}
Q = Q_n + Q_p =\sqrt{\frac{16\pi}5} \Big(\frac3{4\pi} AR^2\beta_2\Big).
\end{equation}
The root mean square (rms) matter radius is defined as
\begin{equation}
\langle r_m^2\rangle=\frac{1}{A}\int\rho(r_{\perp},z) r^2d\tau,
\end{equation}
where $A$ is the mass number, and $\rho(r_{\perp},z)$ is the axially deformed density. The total binding energy and other observable are also attained by using the standard relations, given in Ref. \cite{Pann87}. Here, we have used the NL3$^*$ parameter set with the constant gap BCS approach in the present analysis. More details on BCS pairing and predictive power of RMF for NL3 and NL3$^*$ can be found in Refs. \cite{lala97,Patr09,Niks11,Kara10}.

\begin{figure}[h]
\begin{center}
\includegraphics[width=100mm,height=80mm,scale=1.5]{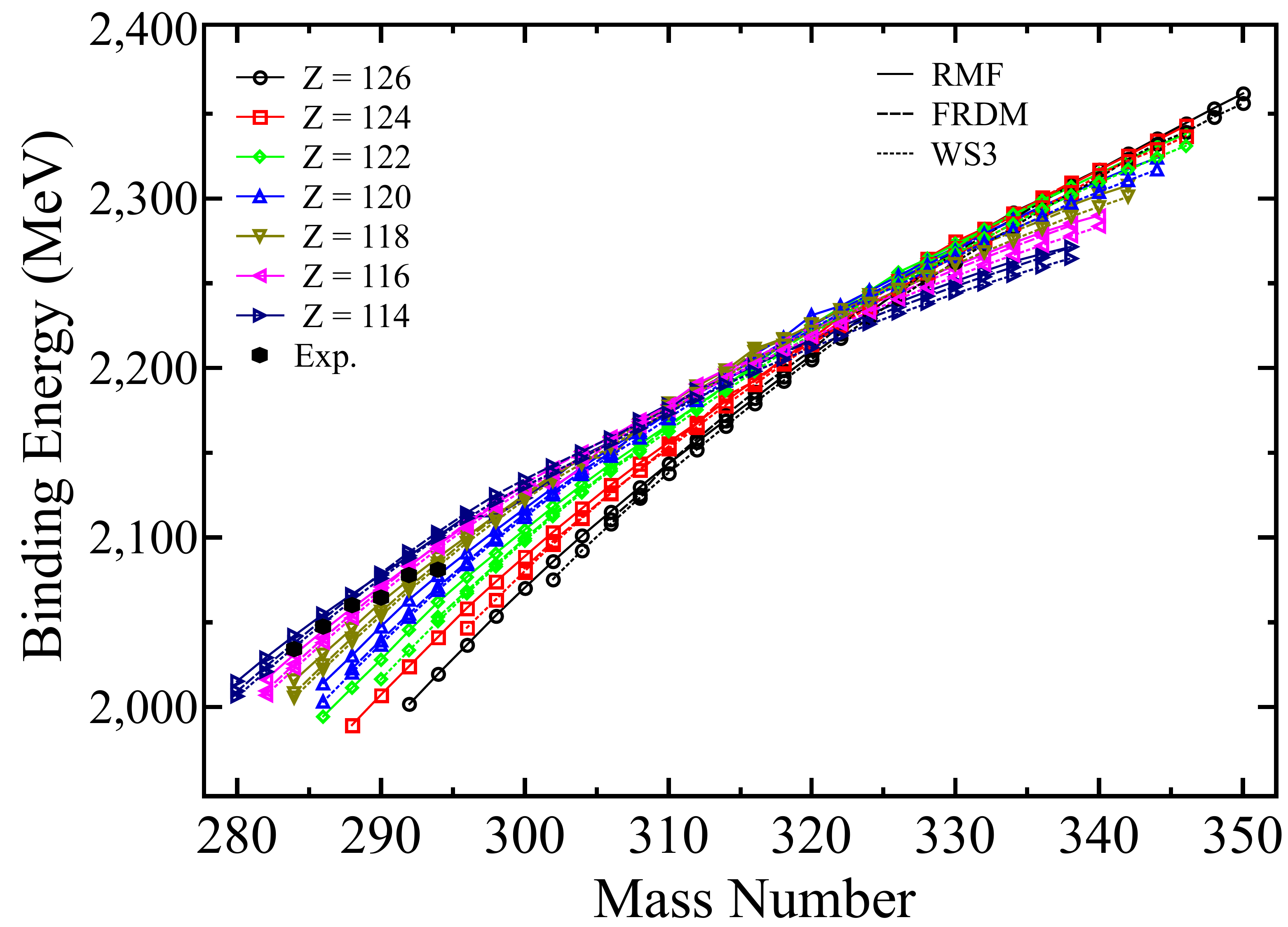}
\vspace{0.05cm}
\caption{\label{fig8} (Color online) The estimated Binding energy of isotopic chain of nuclei from Z = 114 to 126 nuclei by using relativistic mean-field approach with NL3$^*$ parameter (solid line) are given along with the FRDM predictions (dashed line) \cite{moll16}, WS3 predictions (dotted line) \cite{liu11} and the experimental data (dotted symbol) \cite{wang17} for comparison.}
\end{center}
\end{figure}
\begin{figure}[h]
\begin{center}
\includegraphics[width=150mm,height=110mm,scale=1.5]{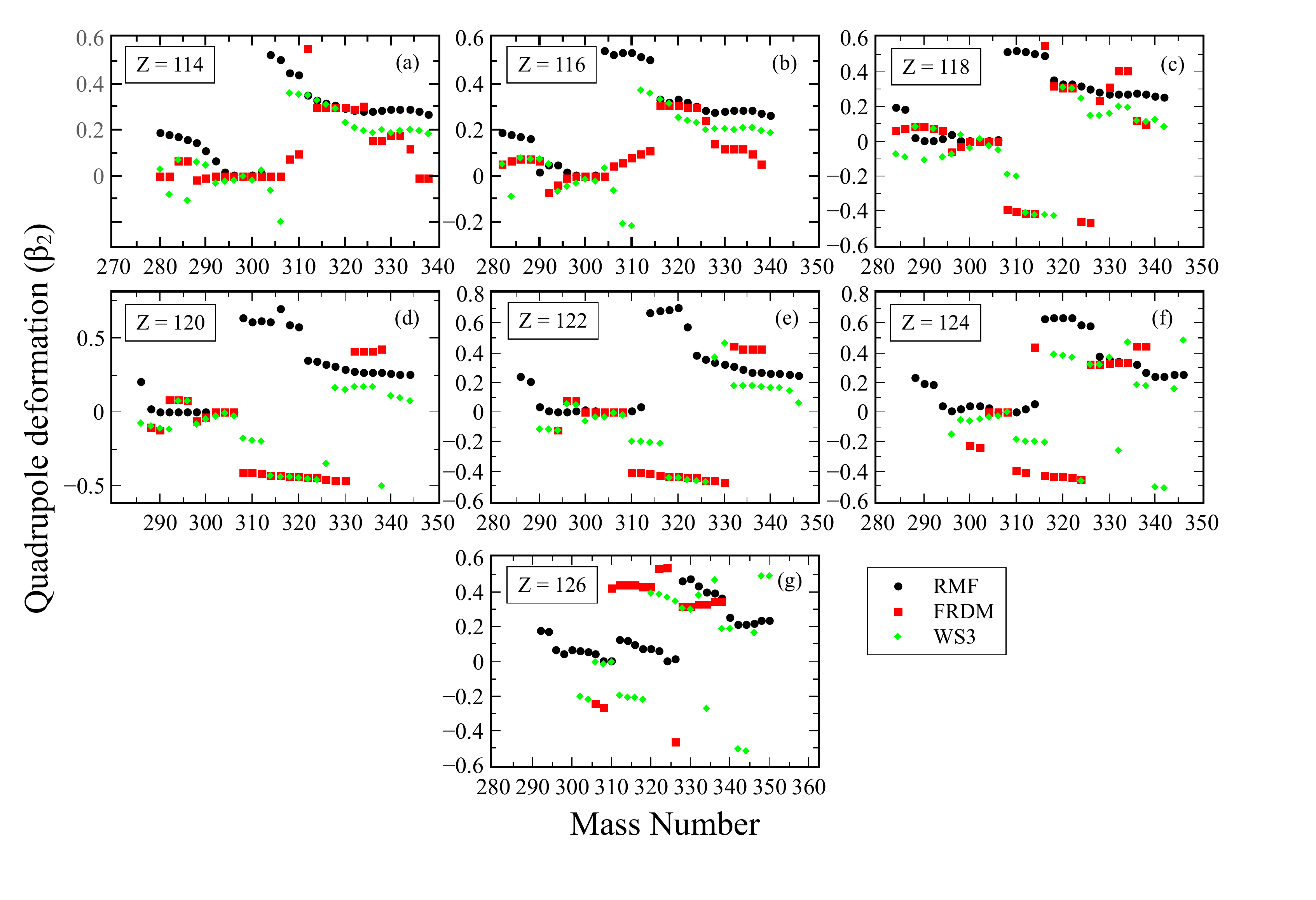}
\vspace{0.05cm}
\caption{\label{fig7} (Color online) The quadrupole deformation parameter ($\beta_2$)
for the isotopic chain of Z = 114 to 126 nuclei by using relativistic mean-field approach with NL3$^*$ are given along with the FRDM predictions \cite{moll16} and Global Nuclear Mass Model (WS3) \cite{liu11} predictions for comparison.}
\end{center}
\end{figure}
\begin{figure}[h]
\begin{center}
\includegraphics[width=100mm,height=80mm,scale=1.5]{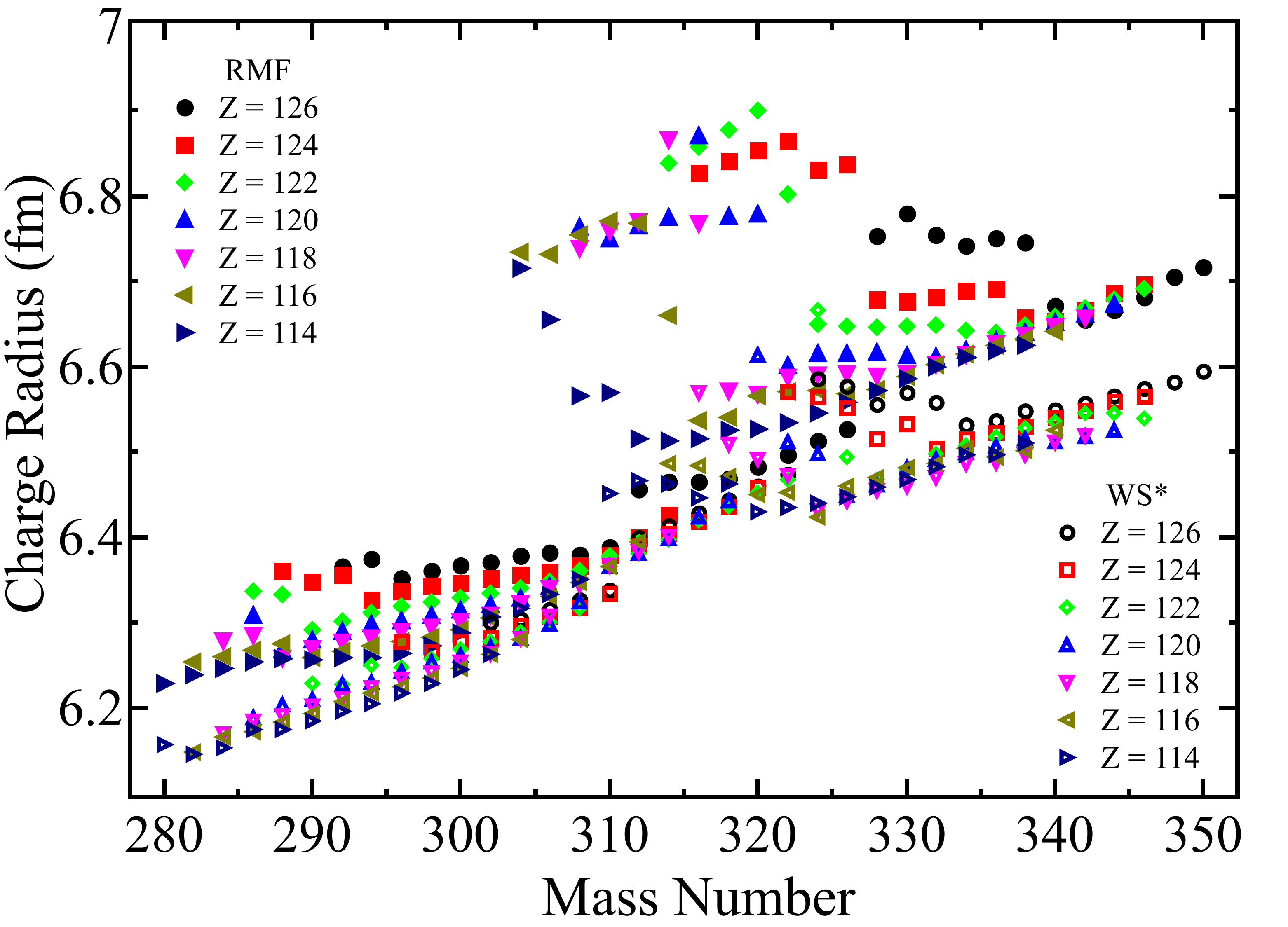}
\vspace{0.05cm}
\caption{\label{fig6} (Color online) The rms charge radius for the isotopic chain of Z = 114 to 126 nuclei by using relativistic mean-field approach with NL3$^*$ are given along with Global Nuclear Mass Model (WS$^*$) predictions \cite{wang10} for comparison.}
\end{center}
\end{figure}

\section{Calculation and Discussions}
\label{result}\noindent
The main aim of the nuclear physicists is to determine the exact predictions for the nuclear masses over the entire nuclear chart because nuclear binding energy provides the primary tool to find the islands of stability and the magic numbers \cite{lunn03}. Hence, finding a convergent solution for the ground and intrinsic excited state for the superheavy region is crucial. The relativistic mean-field equations are solved self-consistently by taking different inputs of the initial deformation $\beta_0$ \cite{Ring96,lala09,lala99}. The convergence of the ground state solutions for this considered superheavy island desire major shells for fermions and bosons is $N_F$ = $N_B$ = 20. The number of mesh points for Gauss-Hermite and Gauss-Laguerre integration is 20 and 24, respectively. We have employed the widely used recently developed non-linear NL3$^*$ parameter sets for the present analysis. It is worth mentioning that the NL3$^*$ \cite{lala09} parameter set is refitted version of the NL3 \cite{lala97} where the exotic nature of the drip-line and superheavy region of the nuclear landscape is considered. Hence, it will be interesting to examine the applicability of relativistic mean-field for NL3$^*$ parameter set and analyze its efficiency in predicting the bulk properties of the nuclei in the superheavy island. Further, the study also provides the relative dependency of the force parameter in the structural properties of superheavy nuclei. In the present work, there is the study about the bulk nuclear properties like binding energy (BE), quadrupole deformation parameter ($\beta_2$), root-mean-square charge radius, and the nuclear decay properties of $even-even$ nuclei with $Z$ = 114 to 126 having the mass number in the range $282\le A \le 350$. The ground state properties as well as the decay properties ($Q_\alpha$ values and  $\alpha$-decay half-lives ($T_{1/2}^{\alpha}$)) of SHN are discussed in the first and second subsections respectively.

\subsection{Bulk nuclear properties of SHN}
\noindent
{\bf Binding energy:} The binding energy is calculated for the superheavy nuclei of atomic number $Z$ = 114 to 126 with a mass number of $280\le A \le 350$ with the RMF model for NL3$^*$ parameter. The RMF results are given along with FRDM predictions  \cite{moll16}, Global Nuclear Mass Model (WS3) predictions \cite{liu11}, and the experimental data \cite{wang17}. It can be observed from Fig. \ref{fig8} that the  RMF results show a reasonably good agreement with the FRDM, WS3 predictions, and the available experimental data. It is also clear from the figure that binding energy increases gradually from $Z$ = 114 to 126 with mass number. We have also calculated the standard deviation of the results obtained from RMF for NL3$^*$,  FRDM, and WS3 predictions with respect to the available experimental data. Quantitatively, the standard deviation of NL3$^*$ parameter set w.r.t available experimental data is 0.9734. In contrast, we find a lower value of standard deviation for FRDM (0.6007) and WS3 (0.3827) compared to NL3$^*$ case. The higher value of standard deviation is an acceptable issue in the self-consistent microscopic calculation as compared to macro-microscopic predictions \cite{Sero86,lala09,lala99}. \\ \\
\textbf{Quadrupole deformation and Charge radius:}The quadrupole deformation parameter provides the shape of nuclei in the ground and intrinsic excited state. In Fig. \ref{fig7}, we have shown the calculated quadrupole deformation parameter for NL3$^*$ parameter set along with the  FRDM \cite{moll16} and WS3 predictions \cite{liu11}. From the figure, one can find the RMF (NL3$^*$) results differ from FRDM \cite{moll16} and WS3 predictions \cite{liu11} for some mass regions. For example, it is estimated from the RMF calculations for NL3$^*$ parameter that all the isotopes of superheavy elements (SHE) in the region $280\le A \le 350$ shows the prolate structure. Although, in some mass regions, we obtained highly deformed prolate structure in the ground state configuration.  In other words, a shape transition was observed for massive isotopes of SHN associated with the atomic number $Z$ = 114 - 126 for NL3$^*$ parameter set. \\ \\
On the other hand, with the increase in mass number, the shape of nuclei changes from prolate to oblate and vice-versa for both FRDM and WS3 predictions. Even though experimental data is not available, it is still more interesting to predict the super-deformed and/or hyperdeformed ground state structure of the SHN within the microscopic models \cite{Bhuy11,lu16} and reference therein. The corresponding root-mean-square ($rms$) charge radius is calculated for superheavy nuclei using RMF formalism for NL3$^*$ parameter set. The calculated $rms$ charge radii are given along with the WS$^*$ prediction \cite{wang10}. It can be noticed from Fig. \ref{fig6} that the $rms$ charge radii monotonically increases with the mass number from $A$ = 280 to $A$ = 350. These monotonic patterns in the charge radius can be correlated with the shape transition of the nuclei in their ground state. In the lower mass region,  the RMF values for $rms$ charge radii have a reasonable agreement with the WS$^*$ values. Whereas it can be seen that for heavier masses, the $rms$ charge values for NL3$^*$ are significantly different from WS$^*$ predictions. These differences are also connected with the ground state configuration of the nuclei. For example, in the heavier mass region, the WS$^*$ predict oblate shape, whereas NL3$^*$ have superdeformed and/or hyperdeformed prolate configuration. \\

\begin{table}[h]
\caption{The possible mean deviation in the $Q$-values for RMF (NL3$^*$) with FRDM predictions \cite{moll95}, WS4 predictions \cite{Wang14}, KUTY mass estimates \cite{chow08} and the experimental data \cite{wang12}. See text for more details.}
\renewcommand{\tabcolsep}{1.0cm}
\renewcommand{\arraystretch}{1.25}
\begin{tabular}{c|c|c}
\hline\hline
Data Points$^a$& With-Respect-To & Mean deviation \\
\hline
5    & Expt-RMF   & -0.2392\\
5    & Expt-FRDM  & 0.2472\\
5    & Expt-KUTY  & 0.5502\\
5   & Expt-WS4   & 0.0262\\
170  & FRDM-RMF   & 0.2265\\
107  & FRDM-KUTY  & 0.2267\\
168  & FRDM-WS4   & -0.0455\\
191 & WS4-RMF    & 0.2192\\
107  & WS4-KUTY   & 0.292\\
\hline \hline
\end{tabular}
\label{tab1}
\footnotesize{$^a$Total number of data points taken for the calculation of mean deviation.}
\end{table}
\begin{figure}[h]
\begin{center}
\includegraphics[width=150mm,height=110mm,scale=1.5]{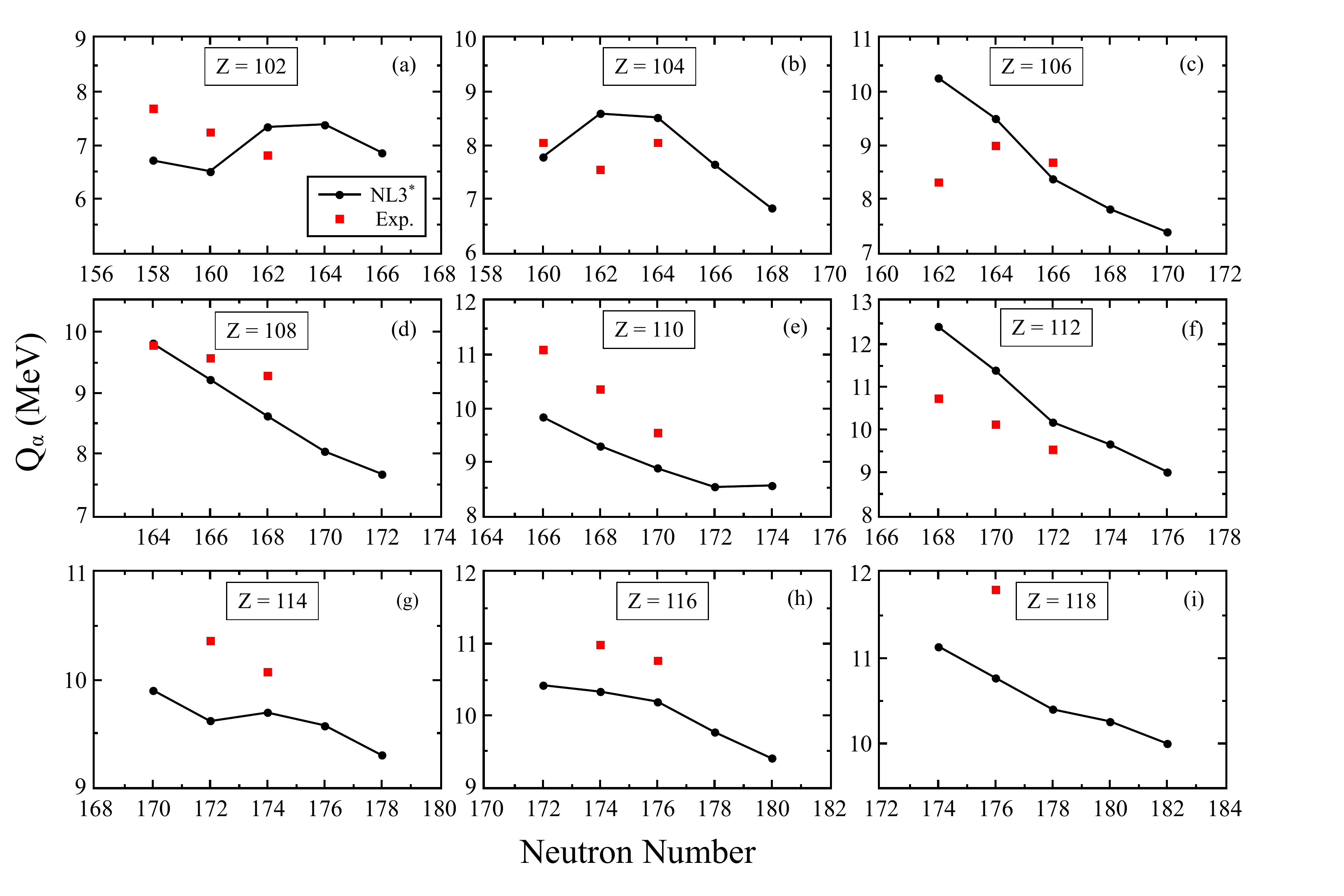}
\caption{\label{figcombo} (Color online) The $\alpha$-decay (Q$_{\alpha}$ energy) for the isotopic chain of Z = 102 to 118 nuclei using RMF (NL3$^*$) are given along with the available Experimental data \cite{wang12}. See text for details.}
\end{center}
\end{figure}
\begin{figure}[h]
\begin{center}
\includegraphics[width=150mm,height=110mm,scale=1.5]{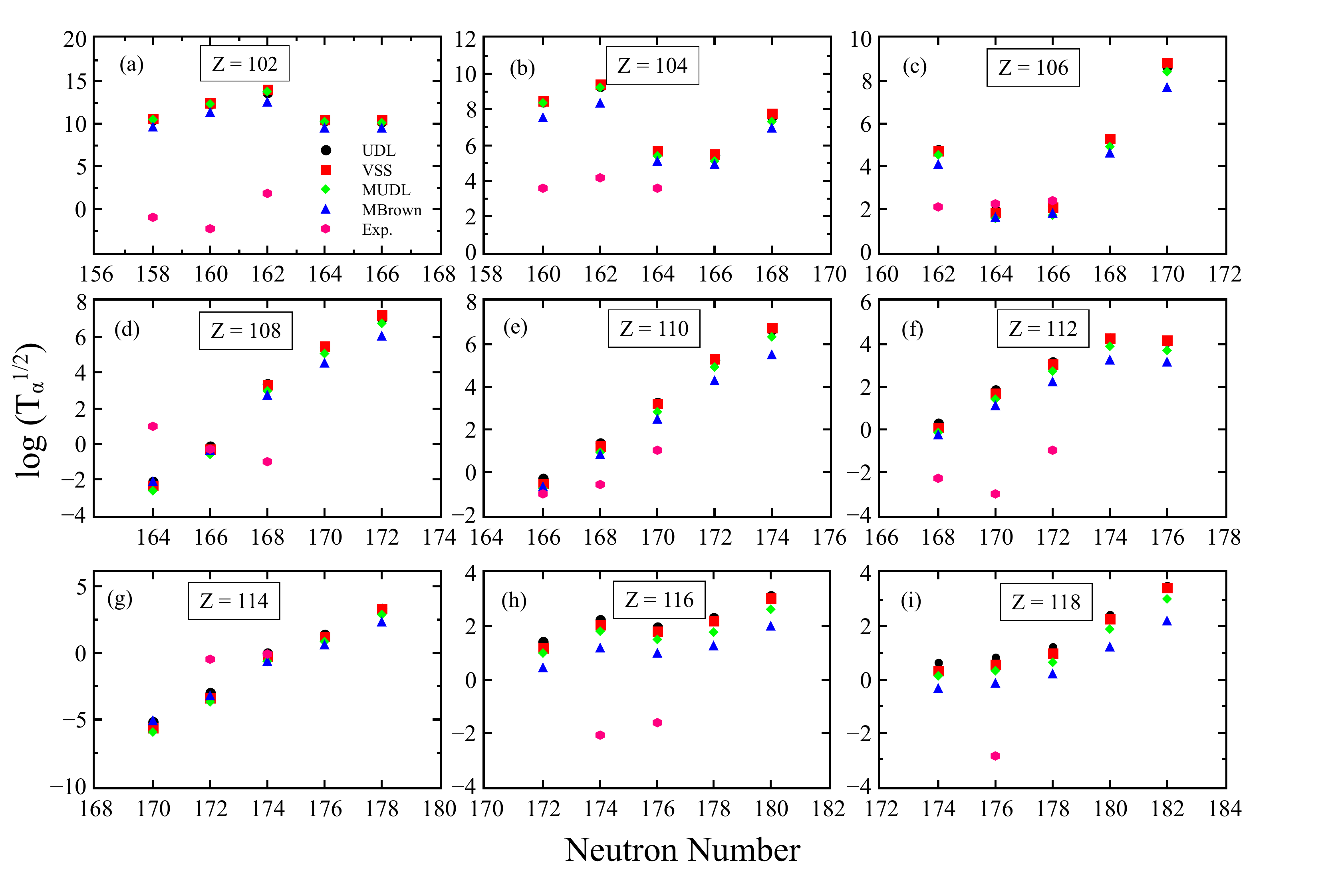}
\vspace{-0.6cm}
\caption{\label{figcom(half)} (Color online) The estimated $\alpha$-decay half-life ($T_{1/2}^{\alpha}$) for the isotopic chain of Z = 102 to 118 nuclei using UDL formula  \cite{manj19}, Viola–Seaborg formula \cite{manj19}, modified UDL formula \cite{akra19} and Modified Brown formula \cite{isma20} for RMF (NL3$^*$) are given along with the FRDM predictions, and available experimental data \cite{Audi17}. See text for details.}
\end{center}
\end{figure}
\begin{figure}[h]
\begin{center}
\includegraphics[width=150mm,height=110mm,scale=1.5]{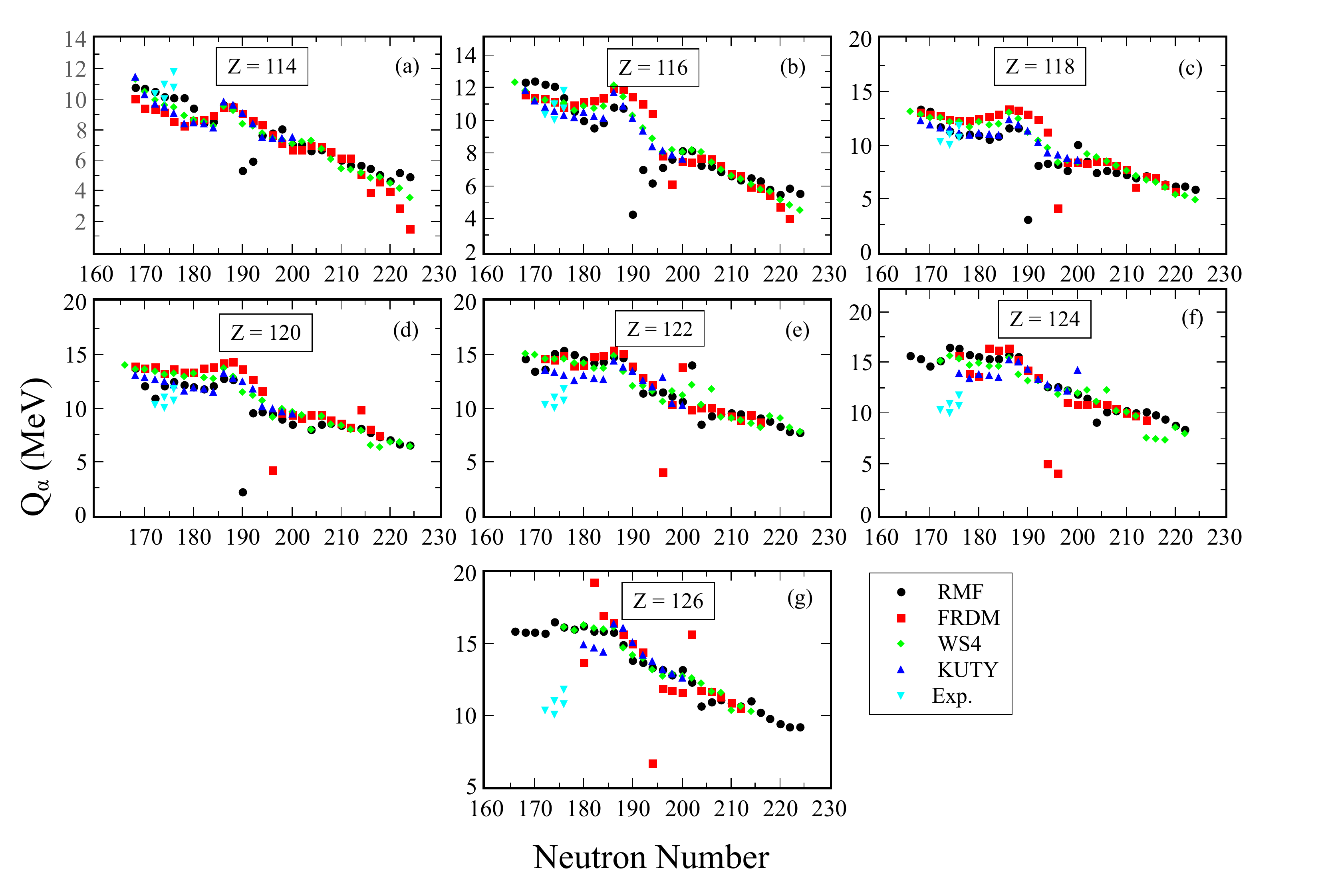}
\caption{\label{fig1} (Color online) The $\alpha$-decay (Q$_{\alpha}$ energy) for the isotopic chain of Z = 114 to 126 nuclei using RMF (NL3$^*$) are given along with the FRDM predictions \cite{moll97}, Global Nuclear Mass Model (WS4) \cite{Wang14} predictions, KUTY mass estimates \cite{chow08} and available experimental data \cite{wang12}. See text for details.}
\end{center}
\end{figure}
\begin{figure}[h]
\begin{center}
\includegraphics[width=150mm,height=110mm,scale=1.5]{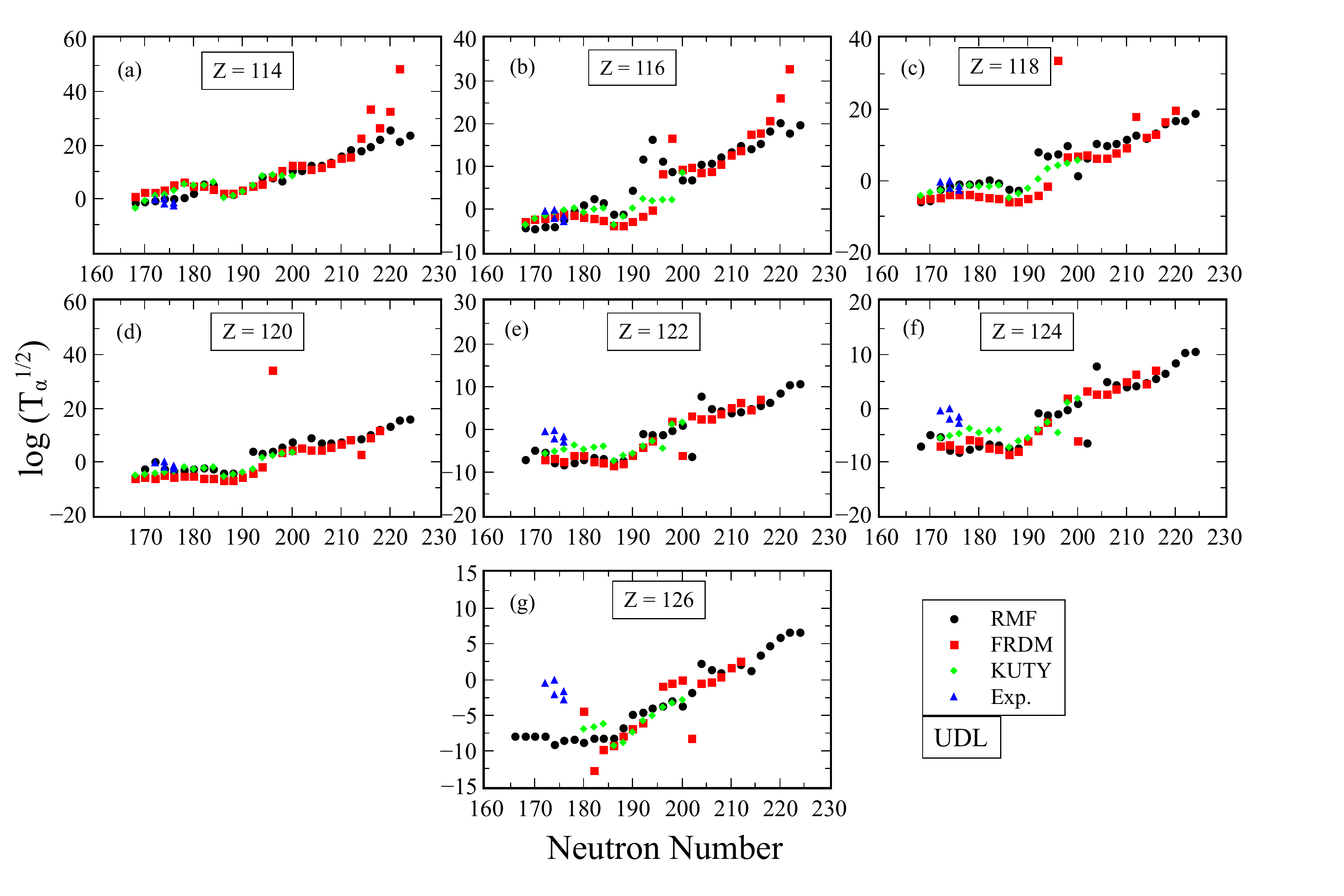}
\vspace{-0.6cm}
\caption{\label{fig2} (Color online) The estimated $\alpha$-decay half-life ($T_{1/2}^{\alpha}$) for the isotopic chain of Z = 114 to 126 nuclei using UDL formula \cite{manj19} for RMF (NL3$^*$) are given along with the FRDM predictions, KUTY mass estimates \cite{chow08} and available experimental data \cite{Audi17}. See text for details.}
\end{center}
\end{figure}
\begin{figure}[h]
\begin{center}
\includegraphics[width=150mm,height=110mm,scale=1.5]{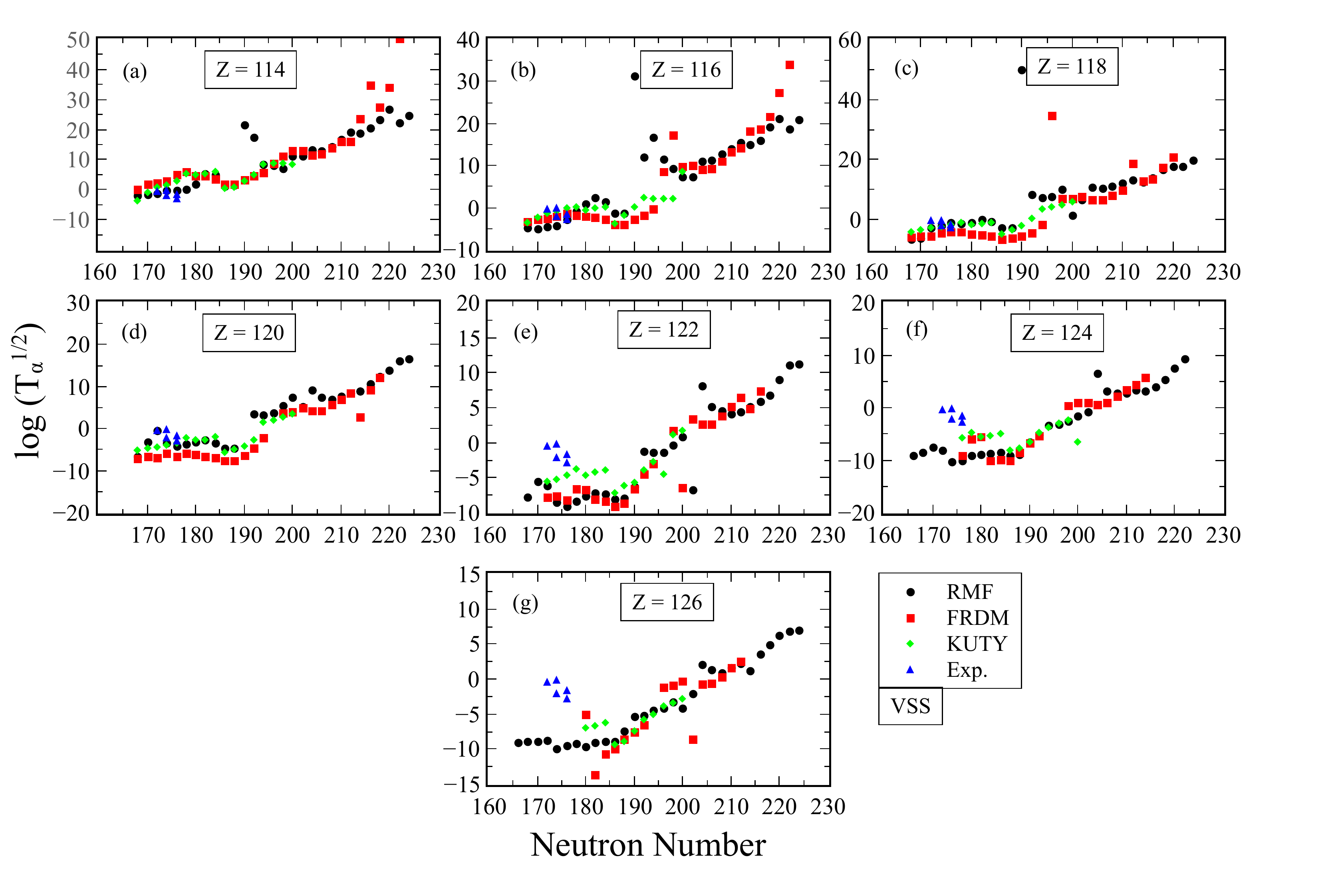}
\caption{\label{fig3} (Color online) Same as Fig. \ref{fig2} but for Viola–Seaborg formula \cite{manj19}. See text for details.}
\end{center}
\end{figure}
\begin{figure}[h]
\begin{center}
\includegraphics[width=150mm,height=110mm,scale=1.5]{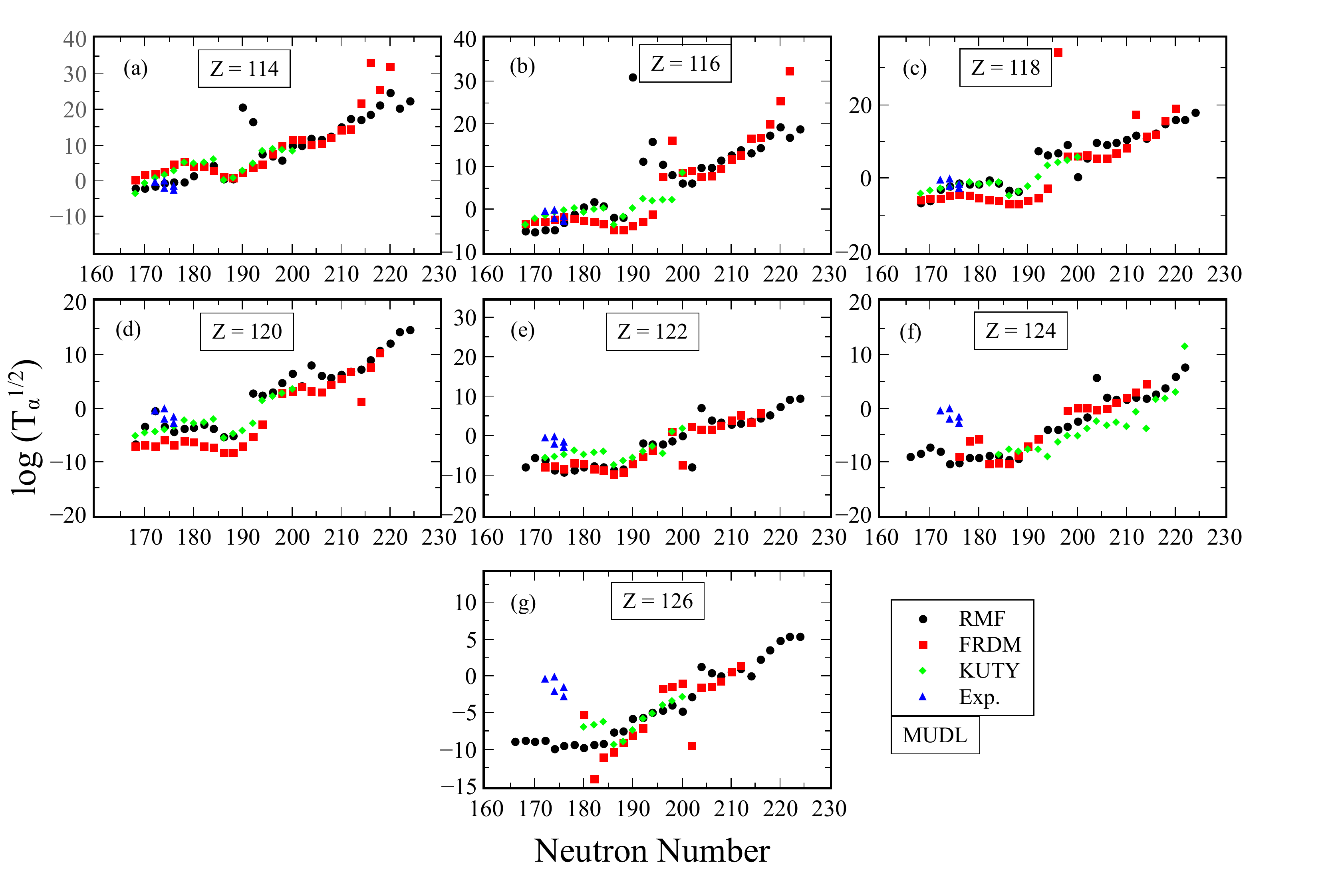}
\vspace{-0.6cm}
\caption{\label{fig4} (Color online) Same as Fig. \ref{fig2} but for modified UDL formula \cite{akra19}. See text for details.}
\end{center}
\end{figure}
\begin{figure}[h]
\begin{center}
\includegraphics[width=150mm,height=110mm,scale=1.5]{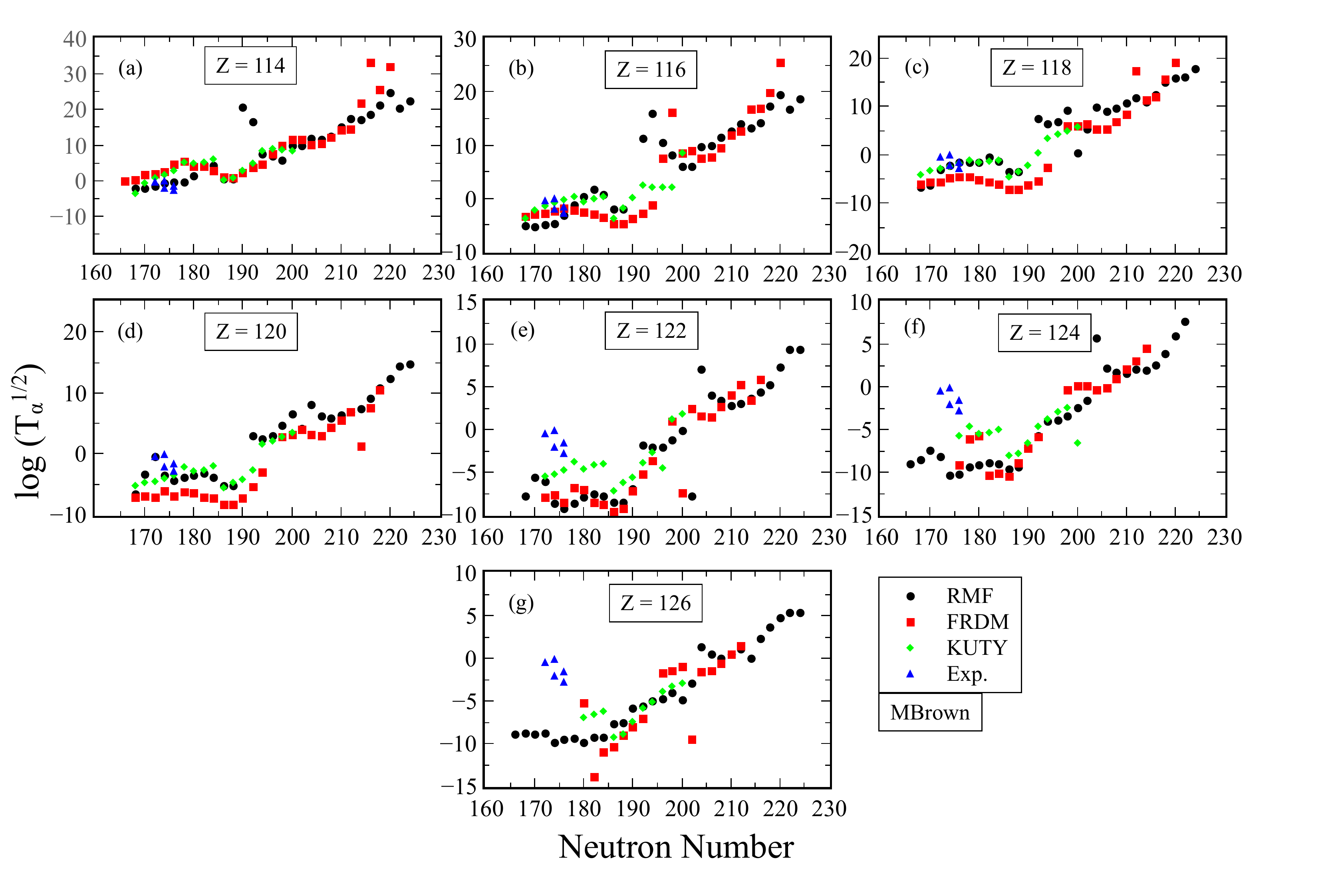}
\vspace{-0.6cm}
\caption{\label{fig5} (Color online) Same as Fig. \ref{fig2} but for Modified Brown formula \cite{isma20}. See text for details.}
\end{center}
\end{figure}
\subsection{Decay Properties of Superheavy Nuclei} 
In our previous works and others, various relativistic parametrizations have been successfully applied to study the ground-state properties such as binding energy, separation energy, quadrupole deformation parameter, $root-mean-square$ rms charge radii, $\alpha$-decay energies, and the alpha decay half-lives of actinides and superheavy nuclei \cite{patr99,bhuy12,Tani20,Bisw21,lala97,Bhuy11,Patr09,Pann87,lala96,Hers06,Zhan06,Sahu11,bhuy15,Bhuy12,bhuy18}. From these analyses, it is found that the results obtained from the NL3$^*$ parameter set provide a reasonable fit to the experimental data. Since experimental data is only available for nuclei up to Z = 118, so, we have further shown the $Q-$values for these nuclei being calculated by using the relativistic mean-field formalism with NL3$^*$ parameter set \cite{Bisw21} and references therein. The $\alpha$-decay half-lives for these nuclei are calculated by using the four different semi-empirical formulae namely UDL, VSS, MUDL, and MBrown (will be discussed in the subsequent subsection). In our previous study also \cite{Bisw21}, we have investigated the $Q-$values and $\alpha$-decay half-lives using NL3$^*$ parameter set with four different formulae, namely, Viola-Seaborg, Alex-Brown, Parkhomenko-Sobiczewski, and Royer.  The results for Q-values and $\alpha-$decay half-lives are shown in Figs. \ref{figcombo}, \ref{figcom(half)}. It can be analyzed from the graphs and also from the previous study \cite{Bisw21} that the results obtained from NL3$^*$ parameter set gives the reliable fit with the experimental data for known superheavy nuclei. Hence, it will be interesting and also crucial to extend the work for the unknown region from Z = 122 to 126 along with two different semi-empirical formulae namely; UDL and MUDL for the calculation of decay half-lives. \\ \\
{\bf The $\alpha$-decay Energy ($Q_\alpha$):} In this subsection, we analyze the $\alpha$-decay energy ($Q_\alpha$) and the alpha decay half-lives ($T_{1/2}^{\alpha}$) of the considered 204 superheavy nuclei associated with various $\alpha$-decay chains. The $\alpha$-decay energy ($Q_\alpha$) plays a significant role in understanding the decay properties and for calculating the $\alpha$-decay half-lives ($T_{1/2}^{\alpha}$) of a nucleus. For example, the $\alpha$-decay half-life of the nuclei having large comparative values in contrast to neighbouring nuclei shows the stability and/or shell closure. The consideration of the $\alpha$-decay of the nuclei determines the island of stability in a superheavy valley. The $Q_\alpha$ energy is calculated using the following relation, $Q_\alpha (N, Z)$ = BE $(N - 2, Z - 2)$ + BE $(2, 2)$ - BE $(N, Z)$. Here, BE $(N, Z)$, BE $(N - 2, Z - 2)$, and BE $(2, 2)$ are the binding energies of the parent, daughter, and $\alpha$ particle (BE = 28.296 MeV), respectively. In the case of FRDM predictions, the $Q_\alpha$ value is calculated by using the mass excess values given in Ref. \cite{moll97}. The calculated $Q_\alpha$ energy values by RMF formalism are compared with the available FRDM data, WS4 predictions \cite{Wang14}, KUTY mass estimates \cite{chow08} and the experimental data \cite{wang12} as shown in Fig. \ref{fig1}. \\ \\
To examine the consistency of the present calculation,  we estimate the mean deviation (MD) in the  $Q_{\alpha}$-values for RMF (NL3$^*$), FRDM, WS4, KUTY mass estimates to the available experimental data. It is worth mentioning that there are only five experimental data available among the 204 considered nuclei. Hence we also estimated the possible mean deviation for RMF (NL3$^*$) with other theoretical models, viz., FRDM, WS4, and KUTY mass estimates. All calculated mean deviations are listed in Table \ref{tab1}. One notices that all the estimated mean deviations (MDs) are in the acceptable range from the table. Further, the comparison MD for RMF (NL3$^*$) with-respect-to experimental data is good compared to FRDM and KUTY mass estimates, whereas WS4 has a lower value. For example, the MD for NL3$^*$ is $\sim -0.23$, and other theoretical models, namely, FRDM,  WS4, KUTY mass estimates, have $\sim 0.25, 0.03,$ and $0.55$, respectively.  This implies that the RMF (NL3$^*$) calculation for  $Q$-values is slightly underestimated to the experimental data, wherever available. But other theoretical models considered here for comparison are slightly overestimated to the experimental data. The relative MDs for NL3$^*$ to experimental data and other theoretical models show that the results are reasonably good agreements with each other.  In other words, the estimated $Q$-values from RMF (NL3$^*$) and other theoretical models apply to the estimation of decay half-life of the unknown region of the superheavy island. Furthermore, the present analysis gives qualitative and quantitative theoretical information on the superheavy island within the microscopic relativistic mean-field model. \\ \\
{\bf The $\alpha$-decay Half-lives:} Geiger and Nuttall predicted the linear relationship between $Q$-value and half-lives of $\alpha$-decay. However, the measured experimental data of parent nuclei emitting $\alpha$-particle in both heavy and superheavy regions with $Z$ = 118 produces several linear segments with varying slopes and intercepts, unlike the Geiger and Nuttall law. To overcome this difficulty, a logarithmic form of experimental $\alpha$-decay half-lives is plotted as a function of different semi-empirical parameters. We have used four different recently developed semi-empirical formulae to calculate the alpha decay half-lives of the superheavy nuclei in this present work. The formulae will be briefly described below: \\
{\bf Formula-1:} Qi {\it et al.} \cite{Qi09} presented a linear universal decay law (UDL) for $\alpha$-decay and cluster decay modes. Two parameters $\chi'$ and $\rho'$ presented in this formula [given in Eq. (\ref{qi-alpha})] depend on the Q-value. The UDL formula relate the half-life of monopole radioactive decay with the Q-value of the emitting particles and the masses and charges of the nuclei involved in the decay. Simply, it is given as,  
\begin{eqnarray}
\log_{10} T_{1/2}^{UDL} &= & aZ_\alpha Z_d \sqrt{\frac{A}{Q_\alpha}} + b\sqrt{AZ_\alpha Z_d(A_d^{1/3} + A_\alpha^{1/3}}) + c \nonumber \\
&& = a\chi' + b\rho' + c. 
\label{qi-alpha}
\end{eqnarray}
Here A is the reduced mass $A = A_dA_\alpha/(A_d + A_\alpha$) and $A_d$, $A_\alpha$ are the mass number of daughter and alpha particle. The adjustable parameter $a$, $b$, and $c$ are 0.3949, -0.3693, and -23.7615 respectively are taken from Ref. \cite{manj19}. \\ \\
{\bf Formula-2:} The Viola and Seaborg (VSS) in 1966 predicted a simple Gamov model-based simple formula for calculating the $\alpha$-decay half-lives. The Viola-Seaborg (VSS) \cite{manj19} relation is given as : 
\begin{eqnarray}
\log_{10} T_{1/2}^{VSS} = (aZ+b)Q^{-1/2}_\alpha+(cZ+d)+h_{log}, 
\label{vss-decay}
\end{eqnarray}
where $a$, $b$, $c$, $d$, and $h_{log}$ are the fitting parameter taken from Sobiczewski \cite{sobi89} having values $a$ = 1.66175, $b$ = -8.5166, $c$ = -0.20228, $d$ = -33.9069 and the hindrance factor $h_{log}$ for the nuclei with unpaired nucleons as,
\begin{eqnarray}
Z= even,  N= even,  h_{log} = 0, \nonumber \\
Z= odd, N= even, h_{log} = 0.772, \nonumber \\
Z= even, N= odd, h_{log} = 1.066, \nonumber \\
Z= odd, N= odd, h_{log} = 1.114.  \nonumber 
\end{eqnarray} \\
{\bf Formula-3:} The universal decay law formula was modified by introducing the nuclear asymmetry term ($I$) to the general procedure. A relation gives the modified universal decay law (MUDL):
\begin{eqnarray}
\log_{10} T_{1/2}^{MUDL} &= & aZ_\alpha Z_d \sqrt{\frac{A}{Q_\alpha}} + b\sqrt{AZ_\alpha Z_d(A_d^{1/3}+ A_\alpha^{1/3}}) \nonumber \\
&& + c + dI + eI^2.
\end{eqnarray}
Here, $I$ represents the nuclear asymmetry term, $I = (N - Z)/A$. And the fitting parameters $a$, $b$, $c$, $d$, and $e$ having values 0.41149, -0.42047, -22.89310, 12.13020, and -44.60575 respectively are taken from ref. \cite{akra19}.\\ \\
{\bf Formula-4:} In 1992, Brown proposed a universal decay law by the linear dependence of half-lives on $Z_d^{0.6}Q_\alpha^{-1/2}$ quantity for even-even parents, where $Z_d$ is the charge number of daughter nuclei. This formula was modified by introducing an additional hindrance term depending on parity. The modified Brown (MBrown) formula \cite{isma20} used here is given as:
\begin{eqnarray}
\log_{10} T_{1/2}^{MBrown} = a(Z-2)^bQ^{-1/2} + c + h^{mB1}.
\end{eqnarray}
The constant parameters $a$, $b$, and $c$ are 13.0705, 0.5182, and -47.8867 respectively and the hindrance term ($h^{mB1}$) depends upon parity is,
\begin{equation}
h^{mB1} = 
\begin{cases}
0   \hspace{0.6cm} for  Z,N  even,\nonumber \\
0.6001 \ for \ Z=odd, N=even,\\
0.4666 \ for \  Z=even, N=odd,\\
0.8200 \ for \ Z,N \ odd.
\end{cases}
\end{equation} \\
In this work, we have systematically analyzed the $\alpha$-decay half-life ($T_{\alpha}^{1/2}$) of $even-even$ 204 isotopes of superheavy nuclei with $Z$ = 114 to 126 by using four semi-empirical formulae, namely, UDL, VSS, MUDL and MBrown as shown in  Fig. \ref{fig2} to \ref{fig5} respectively. To examine the consistency of the alpha decay half-lives obtained from four different formulae, we have calculated the half-lives of the known superheavy nuclei whose experimental data are known. The observed half-lives for the known nuclei are available in Ref. \cite{Audi17}.  It is clear from the figures that there is a fair agreement between the UDL results and the experimental data while the other three empirical formulae (VSS, MUDL, and MBrown) show slightly lower values compared to the experimental data.  \\ \\
From the Figs. \ref{fig2} to \ref{fig5}, one can notice that the isotopes associated with the neutron numbers $N$ = 162, 170, 174, 178, 182 and 184 have longer half-lives as compared to the other isotopes in the series, which further strengthen the predictions in Refs. \cite{bhuy12,zhan05,ghod20}. Moreover, in Ref. \cite{Tani20}, the authors demonstrated that alpha decay half-lives increase with the increase in neutron number and found a neutron shell closure at Z = 120, N = 184, 258. However, their study includes the structure of single-particle states and the nuclear matter properties for superheavy nuclei with Z = 90 - 120. On the other hand, the present study includes the unknown superheavy isotopes covering from neutron deficient to the neutron-rich region of the nuclear chart for  Z = 122, 124, and 126. The calculation mainly involves finding the shell/sub-shell closures from their bulk and decay properties. Four semi-empirical formulae, namely,  UDL, VSS, MUDL, and MBrown are used for the calculation of alpha decay half-lives. Our predictions are compared with FRDM, WS3, WS4  predictions. To find the consistency of these formulae, the standard deviation is also calculated with respect to the other theoretical results. Furthermore, we find a few shell/sub-shell closure with longer half-lives for a few new isotopes. For example, in $^{292-350}$126, the isotopes $^{348,350}$126 ($N$ = 222, 224) have highest half-lives over isotopic chain of Z = 126. Similarly, in $^{290-346}$124, $^{290-346}$122, $^{288-344}$120, $^{286-342}$118, $^{284-340}$116, $^{282-338}$114, the isotopes  $^{346}$124 ($N$ = 222), $^{346}$122 ($N$ = 224),  $^{310}$120 ($N$ = 190), $^{308}$118 ($N$ = 190), $^{306}$116 ($N$ = 190), and $^{334}$114 ($N$ = 220) have relatively longer half-life as compared to other superheavy isotopes excluding the isotopes associated with the above defined neutron magic. From the above analysis, it is found that the $\alpha$-decay half-lives are dependent on the choice of the $\alpha$-decay formula. The bulk properties of these nuclei and the prediction for longer half-lives for a few isotopes are informative for the near future experimental synthesis. \\
\section{Summary and Conclusions}
\label{summary} 
We have explored the nuclear ground-state properties like binding energy, quadrupole deformation parameter($\beta_2$), and root-mean-square charge radius for even-even nuclei for the isotopic chain of $Z$ = 114 to 126 within the relativistic mean-field (RMF) formalism with NL3$^*$ parameter set. The $\alpha-$decay energy and the half-lives are also calculated for the known region $Z$ = 104-118 to examine the predictivity of NL3$^*$ parameter set. The calculated Q$_{\alpha}$-values and their corresponding half-lives are compared with the experimental data and are found to be in reasonably good agreement. Hence, the NL3$^*$ parametrization can be applied for the unknown superheavy region of the nuclear landscape as well. We have extended the study by including the unknown superheavy mass region of $Z$ = 114-126, which cover 204 nos. of isotopes. From the analysis, shape transition is observed from spherical/prolate to the superdeformed ground state while moving from lower to heavier mass isotopes. The $\alpha$-decay energies are calculated from the binding energies of the parent and daughter nuclei by using the RMF (NL3$^*$), and corresponding $\alpha$-decay half-lives are calculated by using four different semi-empirical formulae. The calculated results are compared with macroscopic-microscopic Finite-Range-Droplet-Model (FRDM), Global Nuclear Mass Model (WS3, WS4), Weizsacker-Skyrme mass model (WS$^*$) predictions, and the experimental data, wherever available. The possible mean deviations are also estimated for RMF (NL3$^*$) case with the experimental data and various theoretical predictions. The calculated $Q$-values from RMF (NL3$^*$) are reasonably good agreement with the available experimental data. It is worth mentioning that a few (only five experimental) data are available for the $\alpha$-decay energy for the region of the present study. Hence the comparisons are made for various theoretical predictions, such as FRDM, KUTY mass estimates, and WS4 predictions, and found good agreements. The $\alpha$-decay half-lives are calculated using four different semi-empirical formulae, namely,  universal decay law (UDL), Viola and Seaborg (VSS), modified universal decay law (MUDL), and modified Brown formula (MBrown).  We find consistency and good agreements of the calculated results with FRDM and KUTY mass estimates. Comparing the computed half-life for a particular   $Q$-values for four different formulae, we find the UDL estimate is close to the experimental data, wherever available. From the systematic analysis of these results, it is predicted that the shell/sub-shell closure in terms of longer half-lives appears for the isotopes associated with the neutron number,  $N$ = 162, 170, 174, 178, 182, 184, 190, 220, 222, and 224. The present analysis is informative for the near future experimental synthesis. 

\section*{Acknowledgments}
\noindent
This work has been supported by FOSTECT Project Code: FOSTECT.2019B.04, FAPESP Project Nos. 2017/05660-0, and Board of Research in Nuclear Sciences (BRNS), Department of Atomic Energy (DAE), Govt. of India, Sanction No. 58/14/12/2019-BRNS. 


\end{document}